\documentclass[prd,superscriptaddress,twocolumn,nofootinbib,showpacs]{revtex4-2}

\usepackage{amsfonts,amssymb,amsmath}
\usepackage{color}
\usepackage{graphicx}% Include figure files
\usepackage{dcolumn}% Align table columns on decimal point
\usepackage{bm}% bold math
\usepackage{tikz-feynman}

\newcommand{\B}[1]{{\mathbb #1}}
\newcommand{\C}[1]{{\mathcal #1}}

\newcommand{\BS}[1]{{\boldsymbol #1}}
\newcommand{\BF}[1]{{\mathbf #1}}

\newcommand{\beq}{\begin{equation}}
\newcommand{\eeq}{\end{equation}}
\newcommand{\bea}{\begin{eqnarray}}
\newcommand{\eea}{\end{eqnarray}}

\newcommand{\half}{\frac 12}

\newcommand{\Slash}[1]{{\ooalign{\hfil#1\hfil\crcr\raise.167ex\hbox{/}}}}

\begin{document}

%%%%%%%%%%%%%%%%%%%%%%%%%%%%%%%%%%%%%%%%%%%%%%%%

%\preprint{HIP-2004-**/TH, YITP-2004-**}

%%%%%%%%%%%%%%%%%%%%%%%%%%%%%%%%%%%%%%%%%%%%%%%%

\title{Messenger inflation in gauge mediation and superWIMP dark matter}

\author{Shinsuke Kawai}
%\email{kawai@skku.edu}
\affiliation{Department of Physics, 
Sungkyunkwan University,
Suwon 440-746, Korea}
\author{Nobuchika Okada}
%\email{okadan@ua.edu}
\affiliation{
Department of Physics and Astronomy, 
University of Alabama, 
Tuscaloosa, AL35487, USA} 
\date{\today}% It is always \today, today,

%%%%%%%%%%%%%%%%%%%%%%%%%%%%%%%%%%%%%%%%%%%%%%%%
%%%%% Abstract %%%%%

\begin{abstract}
We discuss phenomenological viability of a novel inflationary model in the minimal gauge mediated supersymmetry breaking scenario.
In this model, cosmic inflation is realized in the flat direction along the messenger supermultiplets and a natural dark matter candidate is the gravitino from the out-of-equilibrium decay of the bino-like neutralino at late times, which is called the superWIMP scenario. 
The produced gravitino is warmish and can have a large free-streaming length; thus the cusp anomaly in the small scale structure formation may be mitigated.
We show that the requirement of the Standard Model Higgs boson mass to be $m_{h^0}=125.1$ GeV gives a relation between the spectrum of the cosmic microwave background and the messenger mass $M$.
We find, for the e-folding number $N_e=60$, the Planck 2018 constraints (TT, TE, EE+lowE+lensing+BK15+BAO, 68\% confidence level) give
$M > 3.64\times 10^7$ GeV.
The gravitino dark matter mass is $m_{3/2} < 5.8$ GeV and the supersymmetry breaking scale $\Lambda$ is found to be in the range
$(1.28-1.33)\times 10^6$ GeV.
%$1.28\times 10^6\;{\rm GeV} < \Lambda < 1.33\times 10^6\;{\rm GeV}$.
Future CMB observation is expected to give tighter constraints on these parameters.
\end{abstract}

%%%%%%%%%%%%%%%%%%%%%%%%%%%%%%%%%%%%%%%%%%%%%%%%

%\pacs{12.60.Jv, 04.65.+e, 98.80.Cq, 98.70.Vc}
\maketitle

%%%%%%%%%%%%%%%%%%%%%%%%%%%%%%%%%%%%%%%%%%%%%%%%
%%%%% Body of the Paper %%%%%

%%%%%%%%%%%%%%%%%%%%%%%%%%%%%%%%%%%%%%%%%%%%%%%%
\section{Introduction}\label{sec:Intro}

The direct detection of gravitational waves by the Laser Interferometer Gravitational-wave Observatory and the discovery of the Higgs boson in the Large Hadron Collider have been major achievements in this century.
For fundamental science, equally important are the null results, or the {\em absence} of discoveries.
Today, we are confronted with several null results contrary to expectations of theorists. 
One of them is the non-detection of cosmic microwave background (CMB) B-mode polarisation of primordial origin.
Since many simple inflationary models predict a large tensor mode perturbation and hence detectably large B-mode polarisation, they are now strongly disfavoured by the observation.
Consequently, realistic inflationary model building requires a more sophisticated framework than, for example, the simple chaotic-type inflation models that were once considered standard. 
Another important null result we encountered concerns supersymmetric particles.
It now seems unlikely that the hierarchy problem is solved by the low scale supersymmetry scenario, as was once hoped.
Nevertheless, supersymmetry itself is an exquisite theoretical framework that resolves many of the shortcomings inherent in the Standard Model (SM), such as the radiative stability of the scalar potential and the natural unification of the gauge couplings. 
It is also well known that supersymmetry is one of the few avenues that the local Lorentz-invariant quantum field theory can be consistently extended \cite{Coleman:1967ad,Haag:1974qh}.
Furthermore, supersymmetry is an essential ingredient when constructing an ultraviolet complete theory, such as string theory.
Thus from a theoretical perspective, it is reasonable to suppose that supersymmetry may well be realized at an energy scale beyond our current experimental reach, rather to dismiss it as an outright irrelevance.

Motivated by the precision measurements of the CMB spectrum, we discuss, in this paper, a scenario of cosmic inflation within the Minimal Supersymmetric Standard Model (MSSM).
In this scenario, inflation is implemented by known component fields, and its prediction is shown to be compatible with current CMB observations.
A notable feature of this scenario is that it predicts gravitino dark matter produced non-thermally at late times, which is called the superWIMP dark matter in the literature \cite{Covi:1999ty,Covi:2001nw,Feng:2003xh,Feng:2004mt}.
As supersymmetry is broken in the real world, any phenomenological model of supersymmetry must include a breaking mechanism.
The gauge-mediated supersymmetry breaking (GMSB) \cite{Dine:1981gu,Nappi:1982hm,AlvarezGaume:1981wy,Dine:1993yw,Dine:1994vc,Dine:1995ag} is a simple, elegant, and highly predictive mechanism of supersymmetry breaking which is not afflicted by the conundrum of flavour-changing neutral currents.
In GMSB, the breaking of supersymmetry is communicated from the hidden sector to the MSSM sector by a pair (or pairs) of the messenger supermultiplets $\Phi$ and $\overline\Phi$ that are charged under the SM gauge group and coupled to a hidden sector field $S$, which is singlet under the SM gauge group.
% \cite{Schmitz:2016kyr,Domcke:2014zqa,Domcke:2017rzu}.
% singlet inflation not favoured by Planck
We investigate cosmic inflation along the flat direction for the messenger fields $\Phi$ and $\overline\Phi$.
This possibility seems relatively unexplored.
We use 125.1 GeV of the Higgs mass as a phenomenological input, and assume that the present dark matter abundance consists of the superWIMP.
Then, this {\em messenger inflation} model predicts a relation between the CMB spectrum and the mass of the messengers. 
%There are rich phenomenological consequences.

In the next section we review the GMSB scenario and explain our notations.
SuperWIMP dark matter, which is one of the key features of this scenario, is discussed in Sec.~\ref{sec:SWDM}.
We describe our inflationary model and its predictions in Sec.~\ref{sec:MessInf}, and discuss constraints on the parameter space in Sec.~\ref{sec:constraints}.
In Sec.~\ref{sec:final} we conclude with some remarks.

%%%%%%%%%%%%%%%%%%%%%%%%%%%%%%%%%%%%%%%%%%%%%%%%
\section{Minimal Gauge mediated supersymmetry breaking}\label{sec:mGMSB}

In GMSB, the messenger fields are vector-like pairs of chiral supermultiplets $D$, $D^c$ and $L$, $\overline L$ transforming under the SM gauge group as in the following table:
\begin{figure}[hbt]
  \begin{tabular}{l|ccc}
    & $SU(3)_c$ & $SU(2)_L$ & $U(1)_Y$\\
    \hline\\
    $D^c$ & ${\BS 3}^*$ & ${\BS 1}$ & $+1/3$\\
    $\overline L$ & ${\BS 1}$ & $ {\BS 2}$ & $-1/2$\\
%    \hline
    $D$ & ${\BS 3}$ & ${\BS 1}$ & $-1/3$\\
    $L$ & ${\BS 1}$ & ${\BS 2}$ & $+1/2$
  \end{tabular}
%  \caption{}
\end{figure}

In order to maintain the gauge coupling unification, it is customary to assume that the messengers come in multiplets
\begin{align}\label{eqn:MessComp}
  \overline\Phi=(D^c, \overline L),\quad \Phi=(D, L),
\end{align}
so that 
$\overline\Phi$ is in ${\BS 5}^*$ and $\Phi$ is in ${\BS 5}$ of a global $SU(5)$ symmetry group.
More generally, the messenger fields may be treated as $N_5$ copies\footnote{
$N_5$ is the Dynkin index of the representation:
$N_5=1$ for ${\BS 5^*+\BS 5}$, $N_5=3$ for ${\BS 1\BS 0^*+\BS 1\BS 0}$,
$N_5=5$ for ${\BS 2\BS 4^*+\BS 2\BS 4}$, etc.
}
of such pairs $(\overline\Phi_I, \Phi_I)$, $I=1,2,\cdots,N_5$.
The superpotential for the GMSB scenario is
\begin{align}\label{eqn:Wmess}
  W= (y S+M)\overline\Phi \Phi + W_{\rm hid}(S),
\end{align}
where $S$ is a SM gauge singlet which represents the dynamics of supersymmetry breaking, $M$ is the mass of the messenger fermions, and $y$ is a dimensionless coupling constant.
The hidden sector superpotential $W_{\rm hid}(S)$ is assumed to give a nonzero F-term for the $S$ field so that the supersymmetry is broken in the vacuum,
\begin{align}
  \langle S\rangle = 0 + \theta^2\;F_S,\qquad
  \langle\Phi\rangle = \langle\overline\Phi\rangle=0.
\end{align}
The messenger mass terms then become
\begin{align}
  {\C L}\supset & -M\Psi_{\overline\Phi}\Psi_\Phi+{\rm h.c}\crcr
  & -\left(
  \begin{array}{cc}
    \overline\Phi\ &\Phi^*
  \end{array}
  \right)
  \left(
  \begin{array}{cc}
    M^2 & yF_S\\ yF_S & M^2
  \end{array}
  \right)
  \left(
  \begin{array}{c}
    \overline\Phi^*\\ \Phi
  \end{array}
  \right),
\end{align}
where $\overline\Phi$ and $\Phi$ represent the scalar components, and $\Psi_{\overline\Phi}$ and $\Psi_\Phi$ are the corresponding fermionic counterparts.
The effect of supersymmetry breaking is manifest as the eigenvalues for the scalar masses are seen to deviate from the fermionic masses.
As the squared scalar masses need be positive, we must have $M^2>yF_S$.
The leading contribution to the gaugino masses arises from 1-loop
\begin{align}\label{eqn:gauginomass}
  M_a = N_5\frac{\alpha_a}{4\pi}\Lambda,
\end{align}
whereas the soft masses for scalars are generated at 2-loop,
\begin{align}\label{eqn:softmass}
  m_i^2 = 2N_5\Lambda^2\sum_{a=1}^3 C_a(R_i)\left(\frac{\alpha_a}{4\pi}\right)^2.
\end{align}
Here we included the multiplicities $N_5$ and introduced
\begin{align}\label{eqn:alphaLambda}
  \alpha_a\equiv\frac{g_a^2}{4\pi}\Lambda,\qquad
%\end{align}
%and
%\begin{align}\label{eqn:Lambda}
  \Lambda\equiv\frac{yF_S}{M}.
\end{align}
The index $a=1,2,3$ stands for $U(1)_Y$, $SU(2)_L$, $SU(3)_c$ and $C_a(R_i)$ are the quadratic Casimir for the representation $R_i$.
%$C_3=4/3$ $C_2=3/4$ $C_1=3y_i^2/20$
The gravitino mass is
\begin{align}\label{eqn:m32}
  m_{3/2}=\frac{F_S}{\sqrt 3 M_{\rm P}},
\end{align}
where $M_{\rm P}=2.44\times 10^{18}\;{\rm GeV}$ is the reduced Planck mass.

The minimal GMSB scenario is highly predictive in the sense that the mass spectrum is determined by a handful of parameters\begin{align}
  N_5,\; M,\; \Lambda,\; \tan\beta,\; {\rm sign}\,\mu. 
\end{align}
One may use the Higgs mass \cite{Zyla:2020zbs} % PDG2020
\begin{align}\label{eqn:h0}
  h_0=125.10\pm 0.14\;{\rm GeV}
\end{align}
to constrain the parameter space.
In Table \ref{table1} we show the prediction of the minimal GMSB scenario for the mass parameters using the {\sc softsusy} 4.1.10 package \cite{Allanach:2001kg}, when $N_5$, $\Lambda$ and $\tan\beta$ are given as the input parameters and $M$ is fixed by the condition \eqref{eqn:h0}.
The sign of $\mu$ is chosen to be positive.
Furthermore, if the gravitino mass is known, from \eqref{eqn:alphaLambda} and \eqref{eqn:m32} we may find\footnote{
This $y$ represents the same degree of freedom as $c_{\rm grav}$ commonly used in the literature.
In \cite{Okada:2012gf,Feng:2012rn} which analyze the superWIMP scenario after the discovery of the Higgs boson, $y$ ($n_{\rm grav}$) is set to unity.
}
\begin{align}\label{eqn:yukawa}
  y=\frac{\Lambda M}{\sqrt 3 M_P m_{3/2}}.
\end{align}
As we see in the next section, the requirement for the superWIMP scenario that the non-thermally generated gravitino comprises the observed abundance of dark matter indeed determines the gravitino mass parameter.
In the rest of this paper, for definiteness we choose $\mu>0$ and consider the minimal GMSB with the superWIMP dark matter scenario, where the prediction is entirely given by $N_5$, $M$ and $\tan\beta$.

%%% TABLE 1 %%%
\begin{table*}
\caption{\label{table1}
The mass spectra in the units of GeV and the CMB parameters $n_s$ and $r$.
We used {\sc softsusy} 4.1.10 \cite{Allanach:2001kg}.
The three parameters $N_5$, $M$, $\tan\beta$ are the input, and the remaining mGMSB parameters $\Lambda$ is fixed by the Higgs mass $m_h = 125.1$ GeV.
We assume $\mu>0$.
The gravitino mass $m_{3/2}$ is determined by \eqref{eqn:Omega32} with $\Omega_{3/2}h^2 = 0.120$ in the superWIMP scenario.
The $n_s$ and $r$ values on the first column are for the minimally coupled case ($M=1.87\times 10^7$ for $N_e=50$, $M=1.46\times 10^7$ for $N_e=60$).
}
\begin{ruledtabular}
\begin{tabular}{c|c|c|c|c||c}
%\hline
$\tan\beta$ & \multicolumn{4}{c||}{10} & 45 \\
\hline
$N_5$ & \multicolumn{4}{c||}{1} & 5 \\
\hline
$M$ & $10^7$ & $10^8$ & $10^9$ & $10^{13}$ & $10^{10}$ \\
$\Lambda$&$1.262\times 10^6$&$1.300\times 10^6$&$1.318\times 10^6$&$1.287\times 10^6$ & $3.157\times 10^5$\\
\hline
$h_0$ & \multicolumn{5}{c}{125.1} \\
\hline
$H_0$ & 5924 & 6521 & 6919 & 7456 & 3680 \\
$A_0$ & 5924 & 6521 & 6919 & 7456 & 3680 \\
$H^\pm$ & 5924 & 6522 & 6920 & 7456 & 3681 \\
\hline
$\widetilde g$ & 8087 & 8263 & 8346 & 8087 & 9437 \\
$\widetilde\chi_{1,2}^0$ & 1764, 3290 & 1809, 3370 & 1832, 3407 & 1780, 3296  & 2176, 3893\\
$\widetilde\chi_{3,4}^0$ &  4034, 4039 & 4648, 4650 & 5026, 5028 & 5372,  5374 & 3922, 4026 \\
$\widetilde\chi_{1,2}^\pm$ & 3290, 4039 &  3370, 4651 & 3407, 5028 & 3296, 5374 & 3893, 4027 \\
\hline
$\widetilde u, \widetilde c_{1,2}$ & $1.084\times 10^4$, $1.154\times 10^4$ & $1.080\times 10^4$, $1.157\times 10^4$ & $1.065\times 10^4$, $1.147\times 10^4$& 9664, $1.060\times 10^4$ & 8291, 8763 \\
$\widetilde t_{1,2}$ & 9721, $1.102\times 10^4$ & 9441, $1.094\times 10^4$ & 9097, $1.076\times 10^4$ & 7622, 9702 & 7215, 8045 \\
$\widetilde d, \widetilde s_{1,2}$ & $1.076\times 10^4$, $1.154\times 10^4$ & $1.070\times 10^4$, $1.157\times 10^4$ & $1.054\times 10^4$, $1.147\times 10^4$ & 9460, $1.060\times 10^4$ & 8237, 8763 \\
$\widetilde b_{1,2}$ & $1.073\times 10^4$, $1.101\times 10^4$ & $1.067\times 10^4$, $1.093\times 10^4$ & $1.051\times 10^4$, $1.076\times 10^4$ & 9416, 9700 & 7764, 8044 \\
\hline
$\widetilde\nu_{e,\mu}$ & 4386 & 4627 & 4812 & 5236 & 3224 \\
$\widetilde\nu_\tau$ & 4381 & 4621 & 4804 & 5222 & 3114 \\
$\widetilde e, \widetilde\mu_{1,2}$ & 2344, 4387 & 2547, 4628 & 2734, 4813 & 3479, 5237 & 1766, 3226 \\
$\widetilde\tau_{1,2}$ & 2327, 4382 & 2525, 4622 & 2707, 4805 & 3438, 5223 & 1321, 3118 \\
\hline
$m_{3/2}$ & 6.137 & 5.375 & 4.741 & 2.721 & 376.8\\
\hline
$\Omega_{3/2} h^2$ & \multicolumn{5}{c}{0.120} \\
\hline
$\xi$~~~~~~~~~~~~~~ & (0) & 0.06310 & 1.339 & $2.419\times 10^4$ & 0.007754\\
$n_s$~~~~~~~~~~~~~ & (0.9418) & 0.9606 & 0.9615 & 0.9616 & 0.9565\\
$r$ ($N_e=50$) & (0.3106) & 0.01498 & 0.004713 & 0.004192 & 0.07526\\
\hline
$\xi$~~~~~~~~~~~~~~ & (0) & 0.08371 & 1.612 & $2.883\times 10^4$ & 0.01218\\
$n_s$~~~~~~~~~~~~~ & (0.9512) & 0.9673 & 0.9678 & 0.9678 & 0.9652 \\
$r$ ($N_e=60$) & (0.2600) & 0.008777 & 0.003270 & 0.002964 & 0.03858
%\hline
 %%%%%
\end{tabular} 
\end{ruledtabular}
\end{table*}

%%%%%%%%%%%%%%%%%%%%%%%%%%%%%%%%%%%%%%%%%%%%%%%%
\section{SuperWIMP dark matter}\label{sec:SWDM}

In GMSB, the gauge mediation effects dominate over the gravity mediation effects.
Since the latter contribution is of the order of the gravitino mass, we must have
\begin{align}\label{eqn:GMSBcond}
  m_{3/2}=\frac{F_S}{\sqrt 3 M_{\rm P}}\ll\frac{\alpha_a}{4\pi}\Lambda.
\end{align}
The gaugino masses and the soft masses are both of the order of $\frac{\alpha_a}{4\pi}\Lambda$.
Thus the gravitino is the lightest supersymmetric particle (LSP) in GMSB.
The next lightest supersymmetric particle (NLSP) is either a neutralino or a stau. See Table \ref{table1}.

As the neutralinos are unstable, the freeze-out neutralino dark matter scenario that leads to the WIMP miracle in generic supersymmetry breaking models is not applicable to GMSB. 
However, the gravitino produced as a decay product of the NLSP neutralino is stable and hence is a good candidate of dark matter.
In this gravitino dark matter scenario, dubbed the superWIMP scenario \cite{Feng:2003xh,Feng:2004mt}, the number density of the gravitino dark matter particles is the same as that of the thermally produced long-lived neutralino NLSP.
The relic abundance of the gravitino dark matter is then
\begin{align}\label{eqn:Omega32}
  \Omega_{3/2}h^2 = \Omega_{\widetilde\chi_1^0}h^2\times\left(\frac{m_{3/2}}{m_{\widetilde\chi_1^0}}\right) = 0.120,
\end{align}
where we used the center value of the Planck 2018 result \cite{Aghanim:2018eyx}
$\Omega_ch^2 = 0.120\pm 0.001$
for the dark matter abundance.
For the bino-like neutralino $\widetilde\chi_1^0\simeq \widetilde B$, the abundance of the thermally produced relics is approximated by\footnote{
This formula for the freeze-out relic abundance is well known \cite{Kolb:1990vq}.
We checked for our parameter choices that numerical results of the micrOMEGAs package \cite{Belanger:2001fz,Belanger:2004yn} agree within a percent. 
}
\begin{align}\label{eqn:Omegah2}
  \Omega_{\widetilde\chi_1^0} h^2 \simeq\frac{8.7\times 10^{-11}\times (n+1)x_f^{n+1}\;{\rm GeV}^{-2}}{\sqrt{g_*}\sigma_{\widetilde B}},
\end{align}
where $n=1$ for the p-wave and
\begin{align}
  x_f =& X - \left(n+\half\right) \ln X,\label{eqn:xf} \\
  X\simeq & \ln\left[0.19\times (n+1)\frac{g}{\sqrt{g_*}} M_{\rm P} m_{\widetilde\chi_1^0}\sigma_{\widetilde B}\right],\label{eqn:X}
\end{align}
with $g=2$ the helicity degrees of freedom for the NLSP neutralino 
$\widetilde\chi_1^0$ (Majorana particle).
We use the SM value $g_*=106.75$ for the relativistic degrees of freedom, as the thermal production takes place well below the supersymmetry breaking scale.
The pair annihilation cross section of the bino-like neutralino 
$\sigma_{\widetilde B}$ is dominated by the p-wave pair annihilation process through the exchange of the right-handed charged sleptons in the t-channel,

\begin{center}
\feynmandiagram [layered layout, horizontal=i1 to a] {
% Draw the top and bottom lines
i1 [particle=\(\widetilde{\chi}_1^0\)]
-- [boson,plain] a
-- [plain] f1 [particle=\(\ell_R\)],
i2 [particle=\(\widetilde{\chi}_1^0\)]
-- [boson,plain] b
-- [plain] f2 [particle=\(\overline{\ell}_R\)],
% Draw the internal slepton line
{ [same layer] a -- [scalar, edge label=\(\widetilde{\ell}_R\)] b },
};
\end{center}
which, for a pure bino $\widetilde B$ with the right-handed slepton exchange is evaluated as \cite{ArkaniHamed:2006mb} % Well-Tempered Neutralino
\begin{align}
  \sigma_{\widetilde B} = \frac{3g_Y^4 \rho(1+\rho^2)}{2\pi m_{\widetilde e_R}^2 (1+\rho)^4},
\end{align}
%$x\equiv m_{\widetilde\chi_1^0}/T$, 
where $\rho\equiv (m_{\widetilde\chi_1^0}/m_{\widetilde e_R})^2$.
We evaluate the gauge coupling $g_Y$ at the breaking scale using the {\sc softsusy} 4.1.10 package \cite{Allanach:2001kg};
for the minimal GMSB parameters $\tan\beta=10$, $N_5=1$ and $M=10^{7-13}$, it is found to be $g_Y\simeq 0.367$.

\begin{figure}
\includegraphics[width=85mm]{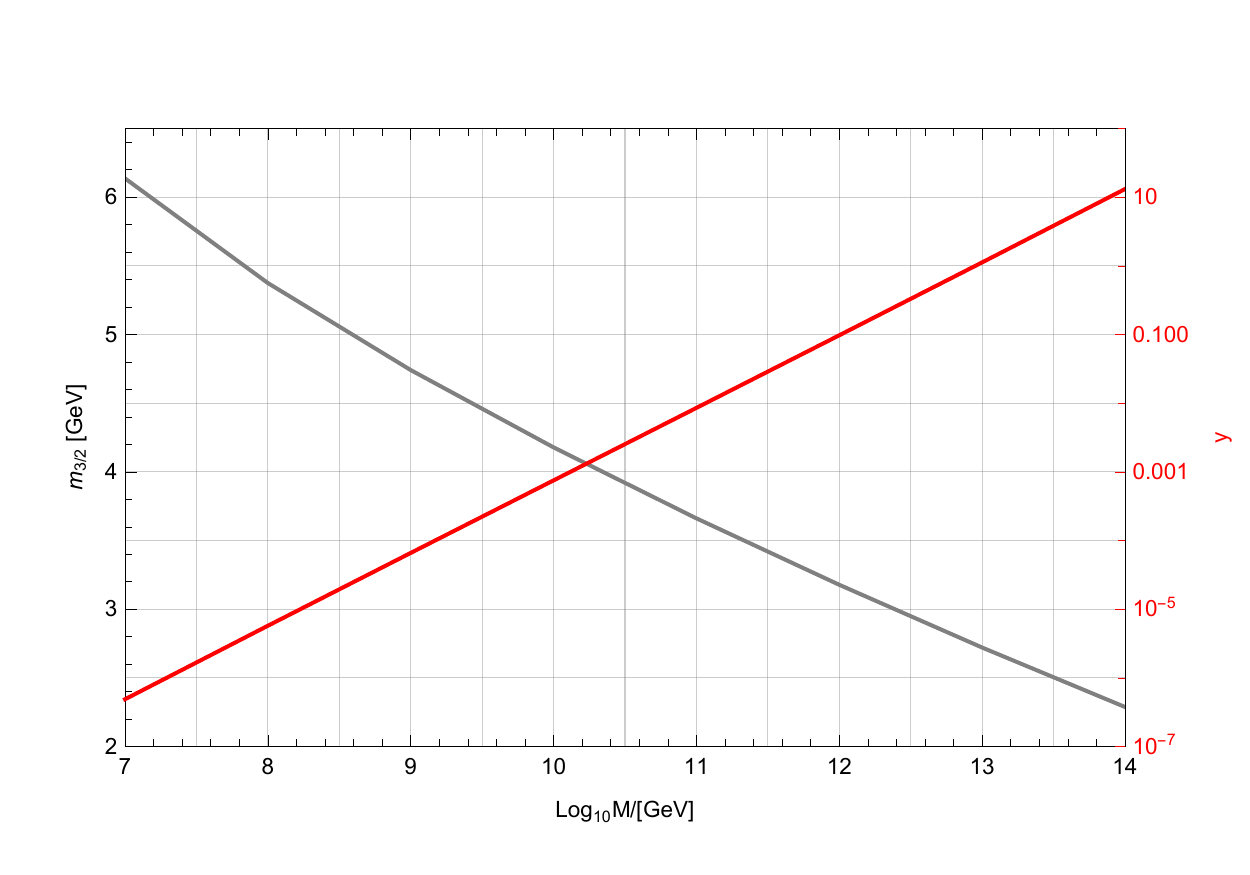}
% Here is how to import EPS art
\caption{
\label{fig:m32andYukawa}
The gravitino mass $m_{3/2}$ (black, scale on the left axis) and the Yukawa coupling $y$ (red, scale on the right axis) as functions of the messenger fermion mass $M$.
We use $N_5 = 1$, $\tan\beta=10$.
The $\Lambda$ parameter is fixed by the condition that the Higgs mass is 125.10 GeV.
}
\end{figure}

The gravitino mass is then determined by the condition \eqref{eqn:Omega32} for the superWIMP scenario, and the Yukawa coupling parameter $y$ is found from \eqref{eqn:yukawa}.
The gravitino mass thus obtained is listed in Table \ref{table1}.
Fig.~\ref{fig:m32andYukawa} shows the gravitino mass $m_{3/2}$ and the Yukawa coupling $y$ as functions of the messenger mass $M$, when $N_5=1$, $\tan\beta=10$ and $\mu>0$.
The gravitino mass stays within a few GeV range, whereas the Yukawa coupling is seen to vary exponentially.
The perturbative limit $y\lesssim {\C O}(1)$ corresponds to $M\lesssim 10^{13}\;{\rm GeV}$.

%%%%%%%%%%%%%%%%%%%%%%%%%%%%%%%%%%%%%%%%%%
%%%%%%%%%%%%%%%%%%%%%%%%%%%%%%%%%%%%%%%%%%

%%%%%%%%%%%%%%%%%%%%%%%%%%%%%%%%%%%%%%%%%%%%%%%%
\section{Messenger inflation}\label{sec:MessInf}

The inflationary model is constructed by embedding the MSSM (including the messenger and the hidden sector) in supergravity.
The part of the Lagrangian pertinent for the inflationary model is
\begin{align}
  {\C L}\supset \int d^4\theta\, \phi^\dagger\phi\,{\C K}
  +\Big\{\int d^2\theta\phi^3 W+{\rm h.c.}\Big\},
\end{align}
where $W$ is the superpotential
%\footnote{While the superpotential \eqref{eqn:Wmess} includes a dimensionful parameter $M$, there is no danger of the tadpole problem since the tadpole effects proportional to the gravitino mass are very small in GMSB.}
given by \eqref{eqn:Wmess} and the K\"ahler potential in the superconformal framework \cite{Kaku:1978nz,Siegel:1978mj,Cremmer:1982en,Ferrara:1983dh,Kugo:1982mr,Kugo:1982cu,Kugo:1983mv}
is here assumed to be noncanonical,
\begin{align}\label{eqn:K}
  {\C K}=&-3M_{\rm P}^2+|\overline\Phi|^2 + |\Phi|^2 + |S|^2\crcr
  &\qquad -\frac 32 \gamma \left(\overline\Phi\Phi+{\rm h.c.}\right)
  -\frac{\zeta}{M_{\rm P}^2}|S|^4,
\end{align}
where $\gamma$ and $\zeta$ are dimensionless parameters.
The mechanism of GMSB discussed in Sec.~\ref{sec:mGMSB} is essentially unaltered by this choice of the K\"ahler potential.

The supergravity embedding is similar to the one used in various supersymmetric extensions of the SM Higgs inflation model \cite{Ferrara:2010in,Arai:2011nq,Pallis:2011gr,Arai:2011aa,Arai:2012em,Arai:2013vaa,Kawai:2014doa,Kawai:2014gqa,Kawai:2015ryj,Leontaris:2016jty,Okada:2017rbf}.
It was discussed in \cite{Einhorn:2009bh} that there is no direct analogue of SM Higgs inflation in the MSSM, whereas in an extended model such as the next-to-minimal supersymmetric Standard Model (NMSSM), a supersymmetric version of the SM Higgs inflation model may be constructed. 
It would be interesting to point out that within GMSB, the MSSM alone can accommodate successful slow-roll inflation of the same type.

%%%%%%%%%%%%%%%%%%%%%%%%%%%%%%%%%%%%%%%%%%
\begin{figure}
\includegraphics[width=24mm]{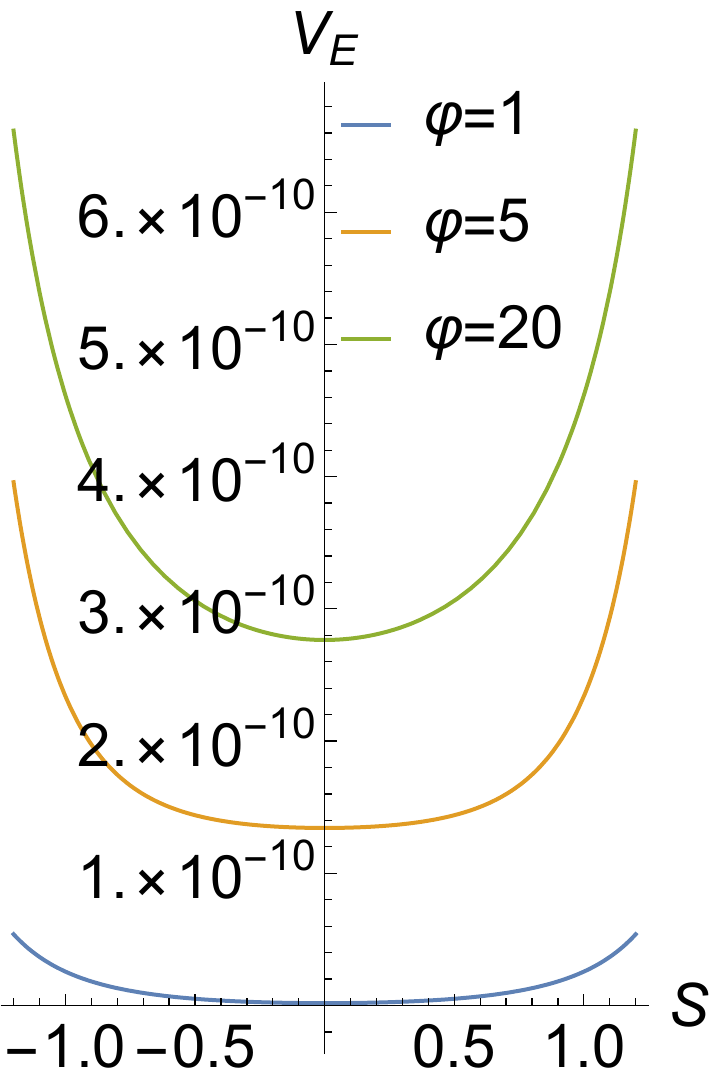}\includegraphics[width=61mm]{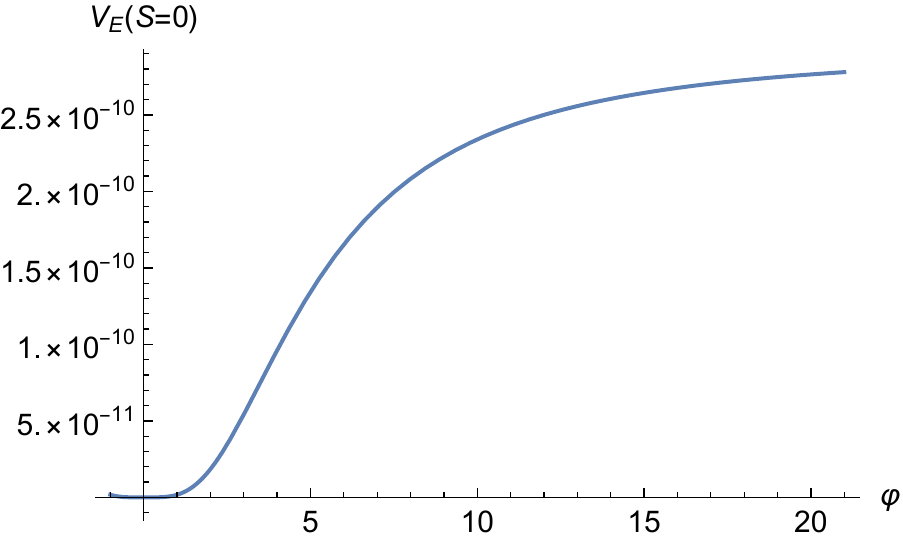}
% Here is how to import EPS art
\caption{
\label{fig:VE}
The scalar potential in the Einstein frame,
for the $\tan\beta=10$, $N_5=1$, $M=10^8$ GeV case of Table \ref{table1}.
The nonminimal coupling is chosen to be $\xi=0.08371$ corresponding to $N_e=60$ e-foldings, and the quartic parameter in the K\"{a}hler potential \eqref{eqn:K} is chosen to be $\zeta=0.1$.
The labels are in the $M_{\rm P}=1$ units.
The left panel shows the potential in the $S$ field direction, with the value of $\varphi$ set at $M_{\rm P}$, $5M_{\rm P}$ and $20M_{\rm P}$.
The potential is seen to be stable at $S=0$.
The right panel show the potential in the $\varphi$ direction, with $S=0$.
}
\end{figure}
%%%%%%%%%%%%%%%%%%%%%%%%%%%%%%%%%%%%%%%%%%

The scalar potential is computed once the superpotential \eqref{eqn:Wmess} and the K\"{a}hler potential \eqref{eqn:K} are given.  
As discussed in \cite{Ferrara:2010yw,Ferrara:2010in}, the instabilities that may exist in the direction of $S$ are controlled by introducing a higher order term in the K\"{a}hler potential, that is, the last term of \eqref{eqn:K}.
The inflationary trajectory is then taken in the flat direction along the messenger multiplets, parametrized by
\begin{align}
  \overline\Phi=\Phi=\half \varphi,\quad\langle S\rangle =0.
\end{align}
In Fig.\ref{fig:VE} we show an example of the scalar potential $V_E(\varphi,S)$ in the Einstein frame, for the $\tan\beta=10$, $N_5=1$, $M=10^8$ GeV and $N_e=60$ case of Table \ref{table1}. 
To be concrete, we use the Polonyi superpotential 
$W_{\rm hid}(S)=\Lambda_S^2 S$, where 
$\Lambda_S\equiv \sqrt{F_S}$, for the hidden sector \eqref{eqn:Wmess} and the quartic parameter is chosen to be 
$\zeta = 0.1$.
When the hidden sector field is stabilized at $S=0$, the scalar part of the action is
\beq\label{eqn:SJ}
S_{\rm scalar}=\int d^4 x\sqrt{-g}\left[
\frac{M_{\rm P}^2+\xi\varphi^2}{2} R
-\half(\partial\varphi)^2
-V_{\rm J}(\varphi)\right],
\eeq
where $R$ is the scalar curvature, $\xi\equiv\frac{\gamma}{4}-\frac 16$ and
\begin{align}\label{eqn:VJ}
V_{\rm J}(\varphi)=&\;
  \frac{y^2}{16}\varphi^4
  +M^2\varphi^2\frac{4M_{\rm P}^2-\left(\frac 16+2\gamma\right)\varphi^2}{8M_{\rm P}^2-\gamma(2-3\gamma)\varphi^2}.
\end{align}
The second term of \eqref{eqn:VJ} is negligible when $y\varphi\gg M$, which is the case of our interest since $M/y\sim 10^{13}\;{\rm GeV}$ (see Fig.\ref{fig:m32andYukawa}) and $\varphi\gtrsim 10^{-2}\,M_{\rm P}$ during inflation.
Thus the effective theory of messenger inflation is the nonminimally coupled $\varphi^4$ inflation model, that includes the minimally coupled $\varphi^4$ model and the SM Higgs inflation model as special cases.
% Conditions for successful slow-roll inflation
% $\varphi^2\lesssim M_{\rm P}^2\lesssim\xi\varphi^2$, 

The prediction of the nonminimal $\varphi^4$ model is understood in the framework of the standard slow roll paradigm \cite{Okada:2010jf}.
Transforming the Jordan frame action \eqref{eqn:SJ} into the Einstein frame by Weyl rescaling of the metric, the field $\varphi$ has a deformed potential
\begin{align}
  V_{\rm E}(\varphi)=\frac{y^2}{16}\frac{\varphi^4}{(M_{\rm P}^2+\xi\varphi^2)^2}.
\end{align}
The field $\varphi$ has a noncanonical kinetic term in the Einstein frame and is related to the canonically normalized field $\widehat\varphi$ by
\begin{align}
  d\widehat\varphi
  =\frac{\sqrt{M_{\rm P}^2+\xi\varphi^2+6\xi^2\varphi^2}}{M_{\rm P}^2+\xi\varphi^2}d\varphi.
\end{align}
One may now introduce the slow roll parameters to analyze the model and obtain the prediction for the cosmological parameters.

The model contains two real parameters $\xi$ and $y$.
We use the normalization of the scalar amplitude\footnote{
In numerics we used the Planck 2018 \cite{Aghanim:2018eyx}
TT, TE, EE +lowE +lensing +BAO value
$As = \exp(3.047)\times 10^{-10}$ at the pivot scale
$k_0 = 0.05\,{\rm Mpc}^{-1}$.}
to determine the nomiminimal coupling $\xi$.
Then, apart from the e-folding number $N_e$, the spectrum is entirely given by a single parameter, which may be chosen to be the Yukawa coupling $y$.

\begin{figure}
\includegraphics[width=85mm]{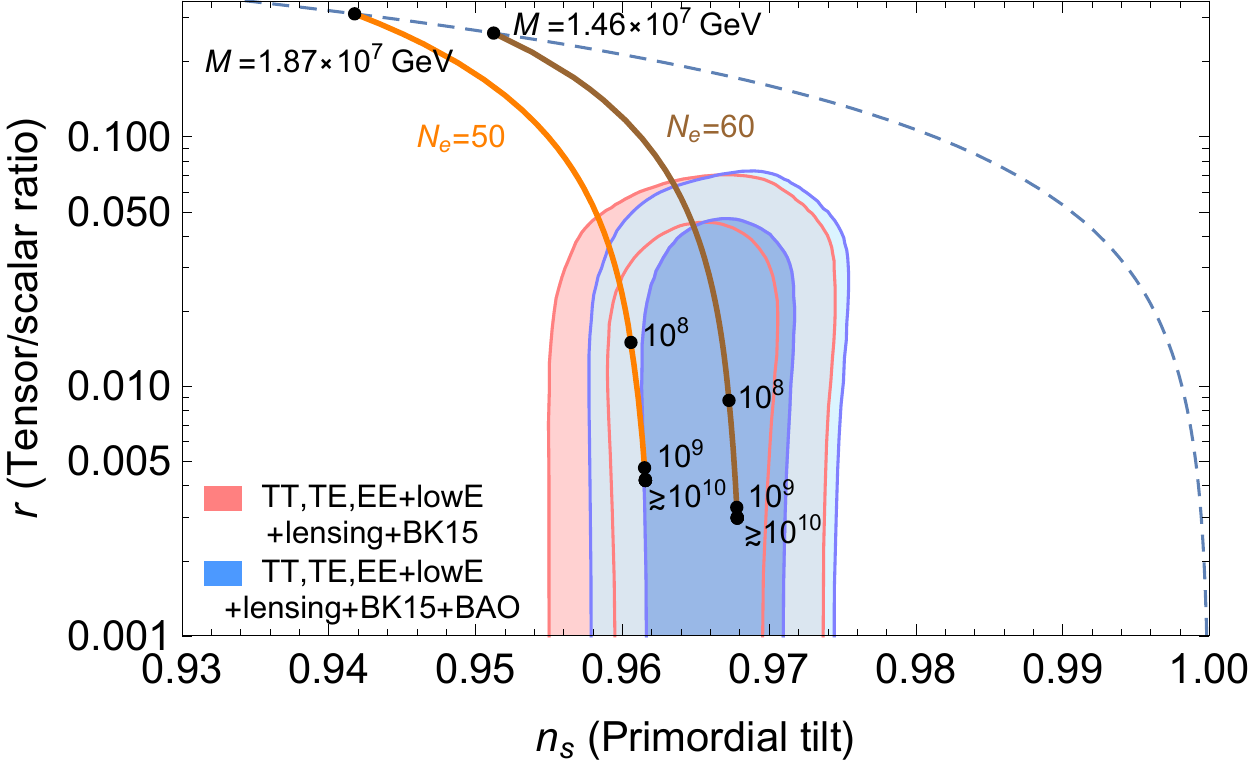}
% Here is how to import EPS art
\caption{
\label{fig:nsrLog}
The prediction of the messenger inflation model for the primordial tilt $n_s$ and the tensor-to-scalar ratio $r$.
The plots for the e-folding number $N_e=50$ and $60$ are shown, with the Planck 2018 1- and 2-$\sigma$ constraints indicated on the background.
The dashed line is $r=\frac{16}{3}(1-n_s)$, corresponding to the minimally coupled ($\xi=0$) case.
}
\end{figure}

In Sec.~\ref{sec:mGMSB} and Sec.~\ref{sec:SWDM} we saw that the minimal GMSB with the successful superWIMP dark matter scenario is controlled by a set of parameters
\begin{align}\label{eqn:params}
  N_5,\quad M,\quad \tan\beta.
\end{align}
The Yukawa coupling \eqref{eqn:yukawa} is also determined by them.
Thus, for a given e-folding number $N_e$, the CMB spectrum of this inflationary model is uniquely determined by the parameter set \eqref{eqn:params}.
Fig.~\ref{fig:nsrLog} shows the primordial tilt $n_s$ and the tensor-to-scalar ratio $r$, when $N_5=1$ and $\tan\beta=10$, as $M$ is varied.
The scalar self coupling for the minimally coupled $\varphi^4$ inflation corresponds to $M=1.87\times 10^7$ GeV for $N_e=50$ and $M=1.46\times 10^7$ GeV for $N_e=60$ and thus, the messenger mass $M$ giving the Planck-normalized scalar amplitude must be larger than these values.

The constraints on the CMB spectrum give the lower bound on the messenger mass $M$. 
We find, from the Planck 2018 results (TT +TE +EE +lowE +lensing +BK15 +BAO) \cite{Akrami:2018odb},
\begin{align}
  M>4.40\times 10^8\; {\rm GeV} & \qquad (\text{68\% C.L.})\\
  M>5.78\times 10^7\; {\rm GeV} & \qquad (\text{95\% C.L.})
\end{align}
for e-folding number $N_e=50$, and
\begin{align}
  M>3.64\times 10^7\; {\rm GeV} & \qquad (\text{68\% C.L.})\\
  M>3.01\times 10^7\; {\rm GeV} & \qquad (\text{95\% C.L.})
\end{align}
for $N_e=60$.
We saw, in Fig.~\ref{fig:m32andYukawa}, that perturbativity $y\lesssim {\C O}(1)$ requires $M\lesssim 10^{13}\;{\rm GeV}$.
In the context of nonminimally coupled $\varphi^4$ inflation, it is argued, e.g. in \cite{Barbon:2009ya,Burgess:2009ea,Hertzberg:2010dc,Lerner:2009na}, that unitarity is violated for too large $\xi$ (see also \cite{Bezrukov:2010jz}, however).
In our case, this gives another upper bound for the messenger mass $M$. 
If the unitarity bound is given by $\xi\lesssim 100$, for example, we must have $M\lesssim 10^{11}$ GeV for both $N_e=50$ and $N_e=60$, which is stronger than the one given by the perturbativity requirement.

%%%%%%%%%%%%%%%%%%%%%%%%%%%%%%%%%%%%%%%%%%
%%%%%%%%%%%%%%%%%%%%%%%%%%%%%%%%%%%%%%%%%%

%%%%%%%%%%%%%%%%%%%%%%%%%%%%%%%%%%%%%%%%%%%%%%%%
\section{Theoretical and observational constraints}\label{sec:constraints}

Let us now discuss further details of the model and its phenomenological consistency.
Our main focus is on the $N_5=1$ and $\tan\beta=10$ case, except in Sec.~\ref{sec:stauNLSP} where the viability of the stau NLSP scenario is discussed.

%%%
\subsection{Reheating temperature}
%%%

Although the messenger fields are coupled to the SM matter fields by the SM gauge interactions, the decay channel of the messenger inflaton is through a Yukawa interaction, if such an interaction exists at all.
Indeed, the $\overline\Phi$ field is in the ${\BS 5}^*$ representation of $SU(5)$, and recalling that this is the same representation as that of the SM fermions (the lepton doublets and the down-type quark singlets) in the Georgi-Glashow $SU(5)$ grand unified theory, a gauge invariant ${\BS 5}^*\cdot{\BS 1\BS 0}\cdot{\BF 5}^*$ Yukawa interaction can exist for $\overline\Phi$.
Using the component fields \eqref{eqn:MessComp}, the superpotential responsible for the decay of the $\overline\Phi$ would be
\begin{align}\label{eqn:Wrh}
  W\supset y_L\overline L H_d e^c + y_D D^c H_d Q,
\end{align}
where $e^c$ and $Q$ are the MSSM fields in ${\BS 1\BS 0}$ and $H_d$ is the down-type Higgs in ${\BF 5}^*$ of $SU(5)$.
Although this Yukawa interaction breaks the ${\B Z}_2$ symmetry that exchanges $\Phi$ and $\overline\Phi$, it has negligible effects on the GMSB mechanism.

The reheating temperature may then be evaluated for the perturbative decay 
$\overline L\to \widetilde H_d\, e^c$ through the $y_L$ interaction\footnote{We neglect nonlinear effects
\cite{Dolgov:1989us,Traschen:1990sw,Kawai:2015lja} for simplicity.}.
The condition that the decay width
\begin{align}
  \Gamma\simeq \frac{y_L^2}{8\pi}M
\end{align}
becomes comparable to the Hubble parameter gives
\begin{align}
  T_{\rm rh}\simeq\frac{y_L}{2\pi}\sqrt{MM_{\rm P}}\left(\frac{45}{2g_*}\right)^{1/4}.
\end{align}
The reheating temperature is thus controlled by the coupling $y_L$.

%%%
\subsection{Thermal gravitino production}
%%%

The gravitino is produced also by the scattering of particles in the thermal plasma during the radiation dominated era after inflation.
Its relic abundance is evaluated as \cite{Bolz:1998ek,Bolz:2000fu,Eberl:2020fml}
\begin{align}
  \Omega_{3/2}^{\rm TP}h^2\sim 0.3 \left(\frac{T_{\rm rh}}{10^{8}\;{\rm GeV}}\right)\left(\frac{1\;{\rm GeV}}{m_{3/2}}\right)\left(\frac{M_3}{1\;{\rm TeV}}\right)^2,
\end{align}
where $M_3$ is the gluino mass.
For the successful superWIMP dark matter scenario, this contribution should be subdominant to the nonthermally produced relic abundance \eqref{eqn:Omega32}.
For $N_5=1$ and $\tan\beta=10$, the gluino mass is $M_3\sim 10\;{\rm TeV}$.
As the gravitino mass range of our interest is $m_{3/2}\sim 1\;{\rm GeV}$, the thermally produced gravitino is negligible when
\begin{align}\label{eqn:TrhUpper}
  T_{\rm rh}\lesssim 10^6\;{\rm GeV}.
\end{align}
On the other hand, the NLSP neutralino of mass $m_{\widetilde\chi_1^0}\sim\text{a few TeV}$ need to be produced from the thermal plasma and thus
\begin{align}\label{eqn:TrhLower}
  T_{\rm rh}\gtrsim\text{a few TeV}.
\end{align}
The reheating temperature of this scenario must satisfy both \eqref{eqn:TrhLower} and \eqref{eqn:TrhUpper}.
For a given messenger mass $M$, the constraints on the reheating temperature give bounds on the Yukawa coupling $y_L$ in \eqref{eqn:Wrh}.
For example, $T_{\rm rh}\sim 10^5\;{\rm GeV}$ is a reasonable value of reheating temperature in our scenario with $M\sim 10^8\;{\rm GeV}$ and $y_L\sim 10^{-8}$.

%%%
\subsection{Big bang nucleosynthesis}
%%%

Another important requirement for the success of the superWIMP scenario is that the big bang nucleosynthesis is not disturbed by the late time decay of the neutralino NLSP \cite{Kawasaki:2004qu}. 
The lifetime of the bino-like neutralino is evaluated for the decay modes $\widetilde B\to \gamma\widetilde G$ as \cite{Feng:2003xh,Feng:2004mt}
\begin{align}
  \tau_{\widetilde B}
  %\simeq\frac{48\pi m_{3/2}^2 M_{\rm P}^2}{m_{\widetilde B}^5}
  \simeq 0.74\times\left(\frac{m_{3/2}}{1\;{\rm GeV}}\right)^2\left(\frac{1\;{\rm TeV}}{m_{\widetilde B}}\right)^5\;\text{sec}.
\end{align}
In our scenario with $N_5=1$ and $\tan\beta=10$, the lifetime of the NLSP neutralino is $\tau_{\widetilde B}<1$ sec.
% $m_{3/2}\sim$ a few GeV and $m_{\widetilde B}\sim$ a few TeV
Thus the NLSP neutralino decays before the big bang nucleosynthesis commences ($\simeq 1$ sec.), leaving the big bang nucleosynthesis intact.

%%%
\subsection{Gravitino free streaming length}
%%%

As the gravitino is much lighter than the NLSP neutralino, the gravitino is energetic when it is produced.
The production occurs at late times, when the Hubble expansion rate is already small and the effect of redshift on the gravitino is not significant.
Thus the nonthermally produced gravitino has a relatively long free streaming length \cite{Borgani:1996ag,Pierpaoli:1997im},
\begin{align}\label{eqn:lambdaFS}
  \lambda_{\rm FS}=&\int_{\tau_{\widetilde B}}^{t_{\rm eq}}\frac{v(t)}{a(t)}dt\simeq 0.18\times \left(\frac{1\;{\rm TeV}}{m_{\widetilde B}}\right)^{3/2}
  \text{Mpc}.
\end{align}
Although the cold dark matter scenario is known to be successful in explaining the structure formation at large scales $\gtrsim 1\;{\rm Mpc}$, recent studies of N-body simulation suggest that the structure formation at small scales favors warm dark matter, with the free streaming length $\lambda_{\rm FS}\simeq 0.1\;{\rm Mpc}$ \cite{Lin:2000qq}.
It can be seen from Table \ref{table1} and \eqref{eqn:lambdaFS} that our model indeed gives $\lambda_{\rm FS}\simeq 0.1\;{\rm Mpc}$, preferred by the small scale structure formation\footnote{The effects of the gravitino on the Hubble tension are discussed recently in \cite{Gu:2020ozv}.}.

%%%
\subsection{Competing dark matter scenarios}
%%%

It is well known that in some parameter region of the minimal GMSB scenario, a stau, rather than a neutralino, can be the NLSP.
An example is shown on the rightmost column in Table~\ref{table1}.
Since we are studying the supersymmetric model, some charged configuration in a flat direction, called a Q-ball, can also be stable and behave as dark matter.
Let us briefly comment of these two possibilities below.

%%%
\subsubsection{Stau NLSP scenario}\label{sec:stauNLSP}
%%%

As the gaugino mass \eqref{eqn:gauginomass} is proportional to $N_5$ and the soft scalar mass \eqref{eqn:softmass} is proportional to $\sqrt{N_5}$, the scalar leptons become relatively lighter than neutralinos when $N_5$ is large.
The lightest stau $\widetilde\tau_1$, instead of the lightest neutralino $\chi_1^0$, may then become the NLSP\footnote{
The stau NLSP scenario may also arise from the effects of hidden sector renormalization \cite{Arai:2010ds,Arai:2010qe}.
}.
An example of the parameters and the mass spectrum giving the 125.1 GeV Higgs boson mass is shown in Table~\ref{table1}.
In this case, the late time decay of the stau NLSP produces the gravitino, which may be considered as superWIMP dark matter.

Similarly to the neutralino NLSP case \eqref{eqn:Omega32}, the relic abundance of the nonthermally produced gravitino is related to that of the stau NLSP $\Omega_{\widetilde\tau_1}h^2$ by
\begin{align}\label{eqn:Omega32stau}
  \Omega_{3/2}h^2=\Omega_{\widetilde\tau_1}h^2\times\left(\frac{m_{3/2}}{m_{\widetilde\tau_1}}\right).
\end{align}
For the (purely right handed) stau, the dominant channels of annihilation are through the electromagnetic interaction into two photons and the t-channel annihilation into two taus with bino exchange:
\begin{center}
  \begin{tikzpicture}
    \begin{feynman}%[small]
      \vertex (m) at (0, 0);
      \vertex (a) at (-1,-0.5) {$\widetilde\tau_1$};
      \vertex (b) at ( 1,-0.5) {$\gamma$};
      \vertex (c) at (-1, 0.5) {$\widetilde\tau_1$};
      \vertex (d) at ( 1, 0.5) {$\gamma$};
      \diagram* {
        (d) -- [photon] (m) -- [scalar] (c),
        (b) -- [photon] (m) -- [scalar] (a),
      };
    \end{feynman}
  \end{tikzpicture}
\hspace{40pt}
\feynmandiagram [layered layout, horizontal=i1 to a] {
% Draw the top and bottom lines
i1 [particle=\(\widetilde{\tau}_1\)]
-- [scalar] a
-- [plain] f1 [particle=\(\tau\)],
i2 [particle=\(\widetilde{\tau}_1\)]
-- [scalar] b
-- [plain] f2 [particle=\(\tau\)],
% Draw the internal bino line
{ [same layer] a -- [boson,plain, edge label=\(\widetilde{B}\)] b },
};
\end{center}
of which the latter contribution is subdominant except when the neutralino mass $m_{\widetilde B}$ is very close to the stau mass $m_{\widetilde\tau_1}$\cite{Asaka:2000zh}. 
For the first diagram, the annihilation cross section is $\sigma_{\widetilde\tau_1}=4\pi\alpha_{\rm em}^2/m_{\widetilde\tau_1}^2$, where $\alpha_{\rm em}=g_{\rm em}^2/4\pi$ and $g_{\rm em}$ is the electromagnetic coupling constant.
The stau NLSP abundance $\Omega_{\widetilde\tau_1}h^2$ is evaluated using the contribution from the first diagram, by equations similar to \eqref{eqn:Omegah2}, \eqref{eqn:xf}, \eqref{eqn:X} with 
$n=0$ (s-wave annihilation), $g=2$, and $\sigma_{\widetilde B}$, $m_{\widetilde\chi_1^0}$ replaced by $\sigma_{\widetilde\tau_1}$, $m_{\widetilde\tau_1}$.

As the gravitino relic abundance is given by \eqref{eqn:Omega32stau}, the gravitino mass is determined by the condition $\Omega_{3/2}h^2=0.120$, that is, the gravitino produced by the late time decay of the stau NLSP comprises the total abundance of the present dark matter.
For example, we find $m_{3/2}=376.8$ GeV when $\tan\beta=45$, $N_5=5$ and $M=10^{10}\;{\rm GeV}$, as listed in Table~\ref{table1}.
The gravitino mass of a few hundred GeV and the stau mass of $\sim\;{\rm TeV}$, however, are excluded by the constraints from the big bang nucleosynthesis \cite{Kawasaki:2008qe}.
Although the parameters of the minimal GMSB may be adjusted, the gravitino mass and the stau mass stay in the same order in the stau NLSP case.
It thus seems difficult to construct a stau NLSP version of this inflationary scenario satisfying phenomenological requirements.

%\cite{Okada:2007na}

%%%
\subsubsection{Q-ball dark matter}
%%%

The reheating temperature \eqref{eqn:TrhUpper} is lower than what is required for generic (non-resonant) leptogenesis.
In the supersymmetric Standard Model, it is then natural to suppose that the baryon asymmetry of the Universe is generated along a charged flat direction, i.e. by the Affleck-Dine mechanism \cite{Affleck:1984fy}.
It has been pointed out \cite{Kusenko:1997si} that some charged configurations along a flat direction, called Q-balls, can be stable and hence considered as dark matter.
The conditions for Q-ball dark matter, namely the stability and the correct relic abundance, are studied for example in \cite{Kasuya:2015uka}.
There are known to be two types of Q-balls, called the gauge mediation type and the ``new" type.
For both types, the parameter ranges for the Q-balls to be dark matter are very different from the parameter range of our scenario;
for the reheating temperature \eqref{eqn:TrhUpper} and the gravitino mass of a few GeV, the Q-balls are actually unstable and they cannot remain as dark matter today.

%%%%%%%%%%%%%%%%%%%%%%%%%%%%%%%%%%%%%%%%%%%%%%%%
\section{Final remarks}\label{sec:final}

%predictive
In this paper we discussed the possibilities that the messenger fields in gauge mediation may be identified as the inflaton field for cosmic inflation.
The requirements that the Higgs boson mass is 125.1 GeV and the gravitino dark matter has the correct relic abundance make the minimal GMSB superWIMP scenario extremely predictive. 
In particular, we have shown that there appears a direct relation between the prediction of the CMB spectrum and the messenger mass scale, as indicated in Fig.~\ref{fig:nsrLog}.

%minimalistic (MSSM)
The theoretical framework of this inflationary scenario is minimalistic.
It is the MSSM with the minimal GMSB.
It is amusing to observe that the phenomenologically viable and observationally compatible inflationary scenario can be constructed in such a simple and well known framework.
In a sense, the model presented in this paper may be considered as the simplest supersymmetric analogue of the SM Higgs inflation model. 
Of course, the relatively large 125.1 GeV Higgs boson mass indicates that the supersymmetric particles (except the gravitino) are heavy, so that testing this scenario by collider experiments is difficult.
%LiteBIRD
In cosmology, there are many projects for CMB observation; for example, the CMB B-mode observation by the LiteBIRD satellite experiments \cite{Hazumi:2019lys} targets the tensor-to-scalar ratio of order $\Delta r\lesssim 0.001$.
Fig.~\ref{fig:nsrLog} suggests that our scenario may be falsified by such experiments; otherwise the value of the tensor-to-scalar ratio $r$ will predict the scale of supersymmetry breaking.

%%%%%%%%%%%%%%%%%%%%%%%%%%%%%%%%%%%%%%%%%%%%%%%%
%{\em Acknowledgements.}---
\subsection*{Acknowledgments}
This work was supported in part by the National Research Foundation of Korea Grant-in-Aid for Scientific Research Grant No.
NRF-2018R1D1A1B07051127, the NRF-JSPS Collaboration ``String Axion Cosmology" (S.K) and by the United States Department of Energy Grant No. DE-SC0012447 (N.O.).

\bigskip

%%%%%%%%%%%%%%%%%%%%%%%%%%%%%%%%%%%%%%%%%%%%%%%%
%\appendix
%\section{Appendix A}

%%%%%%%%%%%%%%%%%%%%%%%%%%%%%%%%%%%%%%%%%%%%%%%%

%%%%%%%%%%%%%%%%%%%%%%%%%%%%%%%%%%%%%%%%%%%%%%%% 
%\bibstyle{h-physrev.bst} 
%\providecommand{\href}[2]{#2} 
%\bibliography{C:/Users/Shinsuke/Dropbox/Archive/BibTeX/AdSCFT.bib,C:/Users/Shinsuke/Dropbox/Archive/BibTeX/pheno.bib} 
%\bibliography{/Users/kawai/Dropbox/Archive/BibTeX/AdSCFT.bib,/Users/kawai/Dropbox/Archive/BibTeX/pheno.bib} 

%%%%%%%%%%%%%%%%%%%%%%%%%%%%%%%%%%%%%%%%%%%%%%%%
\end{document}